# Structural phase transitions in the kagome lattice based materials $Cs_{2-x}Rb_xSnCu_3F_{12}$ (x = 0, 0.5, 1.0, 1.5)

L. J. Downie[1], C. Black[1], E. I. Ardashnikova[2], C. C. Tang[3], A. N. Vasiliev[4,5,6], A. N. Golovanov[4], P. S. Berdonosov[2], V. A. Dolgikh[2], P. Lightfoot[1]*

1. School of Chemistry and EaStChem, University of St Andrews, St Andrews, KY16 9ST, UK.
2. Department of Chemistry, Moscow State University, M.V. Lomonosov Moscow State University, 119991, GSP-1, Moscow, Russia
3. Diamond Light Source Ltd., Harwell Science and Innovation Campus, Didcot, OX11 0DE, UK
4. Department of Low Temperature Physics and Superconductivity, Physics Faculty, M.V. Lomonosov Moscow State University, Moscow 119991, Russia.
5. Theoretical Physics and Applied Mathematics Department, Ural Federal University, 620002 Ekaterinburg, Russia.
6. National University of Science and Technology "MISiS", Moscow 119049, Russia.

*Contact author: pl@st-and.ac.uk

**Abstract**

The solid solution $Cs_{2-x}Rb_xSnCu_3F_{12}$ (x = 0, 0.5, 1.0, 1.5) has been investigated crystallographically between 100 and 300 K using synchrotron X-ray powder diffraction and, in the case of x = 0, neutron powder diffraction. For $Cs_2SnCu_3F_{12}$ (x = 0), there is a structural transition from the previously reported room temperature rhombohedral symmetry ($R\bar{3}m$) to monoclinic ($P2_1/n$) symmetry at 170 K. This transformation is repeated for the x = 0.5 composition, but with an increased transition temperature of 250 K. For x = 1.0 the monoclinic phase is found at 300 K, suggesting that the transition temperature is increased even further. For x = 1.5 a different behaviour, more akin to that previously reported for $Rb_2SnCu_3F_{12}$, is found: a single phase transition between rhombohedral symmetry ($R\bar{3}$) and triclinic symmetry ($P\bar{1}$) is found at 280 K. In agreement with previous single crystal studies, $Cs_2SnCu_3F_{12}$ powder exhibits strong antiferromagnetic interactions (Θ ~ -268 K) and long-range magnetic order at $T_N$ ~ 19.3 K. The finite magnetic moment observed for T < $T_N$ might be explained by a Dzyaloshinskii–Moriya interaction, due to the lowering of symmetry from rhombohedral to monoclinic, which was not suggested in the earlier single crystal study.

**Introduction**

The quantum spin liquid state (QSL) is a highly sought after yet rarely observed magnetic state.[1-3] Its origin lies in geometric frustration and is related to the inability of a system to attain long-range magnetic order (LRO) due to competing magnetic interactions. This frustration can lead to a large number of degenerate magnetic ground-states being present[4] and, in cases were the magnetic ion is a $S$ = ½ species, quantum fluctuations dominate, and the system never achieves a unique LRO ground-state.[1] Evidence for such a state is difficult to gather as there is no single defining experiment which can be used to identify a QSL; magnetic susceptibility measurements must often be augmented with solid-state NMR or muon scattering (μ-SR) experiments in order to confirm a lack of spin freezing.[1,5,6] QSL behaviour has been suggested in a number of materials, for example Herbertsmithite[7] and Kapellasite,[8] (both of which contain $Cu^{2+}$ as the $S$ = ½ ion) and DQVOF[5] (which contains $V^{4+}$). All these materials have in common a crystal structure based on the ideal two-dimensional kagome lattice. The kagome lattice is a geometrically frustrated motif composed of vertex-linked co-planar equilateral triangles, with each node representing a magnetic species.

Another kagome system of interest is $Rb_2SnCu_3F_{12}$. This system does not show QSL behaviour but instead adopts a long-range ordered, but non-magnetic ground state called a valence bond solid (VBS).[9] This results from the kagome lattice being structurally imperfect with neighbouring $Cu^{2+}$-$Cu^{2+}$ distances becoming inequivalent (Fig 1). This leads to localised spin pairing in a 'pin-wheel' arrangement along the shorter cation-cation contacts. Another structural curiosity found for $Rb_2SnCu_3F_{12}$ is a re-entrant phase transition, which is found in the powder form but not in single crystals.[10] The sister phase of $Rb_2SnCu_3F_{12}$, $Cs_2SnCu_3F_{12}$, has been reported to exhibit a perfect kagome lattice above 185 K and appears to show long-range magnetic ordering with reducing temperature (20.2 K). $Cs_2SnCu_3F_{12}$ is suggested to undergo a structural transition at 185 K (on reducing temperature) to a phase similar to that found for $Rb_2SnCu_3F_{12}$; why it does not show a VBS state is unknown but it is suggested that a further structural distortion may occur (the authors tentatively propose a hexagonal/rhombohedral symmetry with a doubled *a*-axis relative to the aristotype phase).[11] A third related system, $Cs_2ZrCu_3F_{12}$, is found to undergo long-range magnetic ordering also but possibly of a different origin.[11] More interesting, from a crystallographic point of view, is the fact that $Cs_2ZrCu_3F_{12}$ shows a third type of structural transition behaviour – at 225 K and below a re-ordering of the fluoride ions occurs driven by a change in coordination environment of the larger $Zr^{4+}$ ion (i.e. from octahedral to seven-coordinated Zr).[12] This preference for larger coordination numbers does not occur for $Sn^{4+}$.

Thus, it can be suggested that the family $A_2MCu_3F_{12}$ shows rich structural chemistry at room temperature and below, which depends subtly on the relative sizes of the A and M cations. In the case of $Cs_2SnCu_3F_{12}$, however, the details of the structural transition near 185 K are vague (indeed no full structural model has been published), but even so its behaviour appears markedly different from the more thoroughly studied $Rb_2SnCu_3F_{12}$ and $Cs_2ZrCu_3F_{12}$. These differences in structural, and consequent magnetic, behaviour have prompted the present crystallographic study of $Cs_2SnCu_3F_{12}$ and the solid solution $Cs_{2-x}Rb_xSnCu_3F_{12}$ (x = 0.5, 1.0, 1.5) in order to more precisely clarify the nature of any structural phase transitions.

**Experimental**

For powder crystallographic studies, RbF (Sigma Aldrich, 99.8 %), CsF (Sigma Aldrich, 99.9 %) and $SnF_4$ (Sigma Aldrich) were dried by heating at 390 K under a vacuum of ~ $10^{-4}$ mbar for 24 hours. They were then mixed and ground with $CuF_2$ (Sigma Aldrich, 98 %) under argon in the appropriate molar ratio. The resulting mixtures were then sealed in a gold tube and heated to 873 K at 10 K min$^{-1}$ and held at this temperature for 12 hours before cooling at the same rate under flowing argon. Pure samples were never realised, with the desired phase being at least 75 % by mass of the sample. Minor impurities include $Rb_2SnF_6$, $Cs_2SnF_6$, CuO and $RbCuF_3$.

Samples of $Cs_2SnCu_3F_{12}$ powder for magnetic measurements were synthesised by grinding and pelletising CsF, $SnF_4$ and $CuF_2$ in stoichiometric amounts in an argon filled glove box. The pellet was placed inside a copper tube along with a small crystal of $XeF_2$ (99.5%). The tube, previously sealed at one end, was then sealed by crimping and welding. The now hermetically sealed tube was then heated to 773 K for three days. The resulting sample was then ground and pelletised under argon before being sealed in a copper tube with more $XeF_2$. The heating cycle was then repeated.

Laboratory based powder X-ray diffraction (PXRD), for the purpose of phase identification, was performed on either a PANalytical Empyrean X-ray diffractometer operating in either Bragg-Brentano or transmission geometry or a Stoe Stadi-P operating in transmission mode; both used CuK$_{α1}$ radiation.

Synchrotron X-ray powder diffraction (SXPD) was performed at beamline I11, Diamond Light Source Ltd., UK.[13] Samples were mounted in glass capillaries and measurements conducted in Debye-Scherrer mode. For $Cs_2SnCu_3F_{12}$ data were collected using a multi-analysing crystal detector at 100, 120, 140, 150, 160, 170, 180, 190, 200, 210, 220, 240, 260, 280 and 300 K on heating (220 – 160 K on cooling, also) ("run 1"). A further run was performed with longer collection times ("run 2") and measurements were taken at 300, 200, 190, 180, 170, 165, 160, 150 and 140 K on cooling. For $Cs_{2-x}Rb_xSnCu_3F_{12}$, patterns were collected between 300 and 100 K in 10 K increments utilising a position-sensitive detector. In all cases a minimum of ten minutes was allowed in order for the sample to equilibrate at each temperature.

Neutron powder diffraction (NPD) was performed on $Cs_2SnCu_3F_{12}$ at beamline HRPD, ISIS pulsed neutron and muon source, UK. A Sample of ~ 5 g was mounted in a 5 mm vanadium flat plate can before loading into the diffractometer, which was fitted with a cryostat. Patterns were collected at 300, 165, 150 and 100 K.

Rietveld refinement of powder diffraction patterns was performed using GSAS[14] and the EXPGUI interface.[15] In all cases the background was refined using a shifted Chebyschev function of between 12 and 20 terms. Lattice parameters were also refined along with profile functions, taking account a small amount of preferential orientation. Thermal parameters were refined isotropically, with similar atoms being constrained to be the same in the SXPD data. Atomic coordinates were refined in the case of NPD data and the higher symmetry ($R\bar{3}m$) SXPD data, but these were fixed for the lower symmetry models in the SXPD refinements.

Magnetic measurements were performed using a Quantum Designs PPMS using a VSM add-on unit. Samples were mounted in plastic holders and subject to both zero-field-cooled (ZFC) and field-cooled (FC) measurements between ~ 2 and 300 K.

**Results and Discussion**

For clarity, this section is broken into three parts considering the crystallography of $Cs_2SnCu_3F_{12}$, the crystallography of $Cs_{2-x}Rb_xSnCu_3F_{12}$ and the magnetic properties of $Cs_2SnCu_3F_{12}$ powder.

**Crystallography of $Cs_2SnCu_3F_{12}$**

Both SXPD and NPD patterns at room temperature (ESI, Fig. S1) suggest agreement with the previously reported $R\bar{3}m$ unit cell ($a \approx 7.14$ Å, $c \approx 20.38$ Å) derived from an X-ray single crystal study. (Lattice parameters as a function of temperature are available in the ESI, Tables S2, S3 and S4). Significantly, the atomic coordinates for both the powder (as derived by Rietveld refinement) and the previously reported single crystal diffraction data are in agreement (Table 1).

In SXPD ("run 1") on cooling, a number of peak splittings occur suggesting a change in unit cell symmetry (Fig. 2). This transition is seen to occur at around 170 K – similar to previous reports[11]. Further cooling to 100 K allows the splittings to be clearly resolved, and indexing at this temperature suggests that the new unit cell is monoclinic ($a \approx 7.901$ Å, $b \approx 7.100$ Å, $c \approx 10.597$ Å, $\beta \approx 97.87$ °), with similar parameters being found from the NPD data. The monoclinic cell is related to the rhombohedral cell by the transformation matrix (1/3, 2/3, -1/3), (-1, 0, 0), (2/3, 4/3, 1/3). Rietveld refinement of the NPD data against models of the four possible space groups ($P2_1/n$, $P2_1/m$, $P2/n$, $P2/m$, as generated from the program ISODISTORT[16]) suggest that the $P2_1/n$ model is optimal (ESI, Table S1).

Table 1: Atomic coordinates for $Cs_2SnCu_3F_{12}$ at room temperature (space group $R\bar{3}m$) as derived from single crystal, SXPD and NPD data.

Previously reported single crystal[11]
(a = 7.142(4) Å, c = 20.381(14) Å)

|    | x | y | z |
|----|---|---|---|
| Cs | 0 | 0 | 0.1060(1) |
| Cu | ½ | 0 | 0 |
| Sn | 0 | 0 | ½ |
| F1 | 0.2042(2) | -0.2042(2) | 0.9845(1) |
| F2 | 0.1312(2) | -0.1312(2) | 0.4465(1) |

SXPD Rietveld refinement
(a = 7.131778(10) Å, c = 20.36703(4) Å)

|    | x | y | z |
|----|---|---|---|
| Cs | 0 | 0 | 0.10627(3) |
| Cu | ½ | 0 | 0 |
| Sn | 0 | 0 | ½ |
| F1 | 0.20541(15) | -0.20541(15) | 0.98475(10) |
| F2 | 0.13211(15) | -0.13211(15) | 0.44676(10) |

NPD Rietveld refinement
(a = 7.13155(6) Å, c = 20.3609(3) Å

|    | x | y | z |
|----|---|---|---|
| Cs | 0 | 0 | 0.1057(3) |
| Cu | ½ | 0 | 0 |
| Sn | 0 | 0 | ½ |
| F1 | 0.20419(19) | -0.20419(19) | 0.98422(12) |
| F2 | 0.13008(19) | -0.13009(19) | 0.44697(14) |

Further examination of the phase transition using both SXPD ("run 2") and NPD was performed and these suggest a direct transition between the $R\bar{3}m$ unit cell found at room temperature and the monoclinic $P2_1/n$ unit cell found with lowering temperature. At the first indication of a transition in the SXPD data there is a peak indexable as (1 2 0) and the NPD pattern at 165 K – 5 K below the transition – presents a peak indexable as (0 1 4), also disallowed in a C-centred unit cell. Further to the appearance of peak splittings there is also a continuous change in the gradient of normalised unit cell volume at the transition (Fig. 3), compatible with a $2^{nd}$ order phase transition.

Examination of the resultant low temperature phase from NPD and SXPD shows that the distortion of the kagome lattice is different from that previously suggested for either $Cs_2SnCu_3F_{12}$[11] or $Cs_2ZrCu_3F_{12}$[12] or that of room-temperature $Rb_2SnCu_3F_{12}$. In this case the distortion (Fig. 4) leads to the formation of three different Cu – Cu distances, as opposed to the four found in $Rb_2SnCu_3F_{12}$[9]. The differing nature of these structural phase transitions is presumably related to size effects at both the Rb/Cs and Sn/Zr sites. In the case of $Rb_2SnCu_3F_{12}$ it was found that a symmetry-lowering transition occurs in powder form but not in single crystals.[10]

The motivation for the transition in the present case is unclear, but bond valence sums (BVS)[17] derived from the refinement of NPD data for the room temperature structure suggest that $Sn^{4+}$ is over-bonded (BVS = 4.63) (BVS for all ions in ESI, Table S5, and bond distances also in ESI, Table S6). At 100 K it is found that the Sn – F bond distance has actually increased, suggesting that the distortion is in some way motivated by the need for $Sn^{4+}$ to reduce its BVS (to 4.55).

**Crystallographic studies of the solid-solution $Cs_{2-x}Rb_xSnCu_3F_{12}$ (x = 0.5, 1.0, 1.5)**

$Cs_2SnCu_3F_{12}$ adopts a different room-temperature structure, and shows different phase transition behaviour at low-temperature, when compared to $Rb_2SnCu_3F_{12}$. The solid-solution is therefore of interest.

$Cs_{1.5}Rb_{0.5}SnCu_3F_{12}$ is found to be structurally very similar to $Cs_2SnCu_3F_{12}$ at room-temperature, adopting the unit cell with $R\bar{3}m$ symmetry ($a$ = 7.12113(2) Å, $c$ = 20.35953(10) Å from SXPD data). On cooling, the structural behaviour is similar to that found for $Cs_2SnCu_3F_{12}$; a transition to $P2_1/n$ symmetry is found (lattice parameters as a function of temperature are presented in the ESI, Table S7). The temperature for this transition is found to have increased relative to $Cs_2SnCu_3F_{12}$ and, by observing the quality of the SXPD Rietveld fit for both the rhombohedral and monoclinic models, occurs at ~ 250 K (Fig. 5a). The variation of normalised unit cell volume also exhibits a change in gradient coincident with the structural transition (Fig 5b).

For $CsRbSnCu_3F_{12}$, the assignment of symmetry at room temperature is more challenging – both of the rhombohedral models, previously derived for $Cs_2SnCu_3F_{12}$ and $Rb_2SnCu_3F_{12}$, lead to poor fits (ESI Table S8). The peaks appear to be broadened at room temperature and thus it was decided to commence data analysis at the lowest temperature recorded (100 K). It might be assumed that any splittings due to symmetry-lowering would be maximised at this temperature. Indexing at the lower temperature suggests that the pattern is best fitted by the monoclinic unit cell previously derived for $Cs_2SnCu_3F_{12}$ at low temperature. The symmetry is once again found to be primitive – an observed peak indexable as (1 2 0) eliminates the possibility of a centred cell – and there are no peaks which break the $n$-glide condition. This leaves $P2_1/n$ and $P2/n$, of which $P2_1/n$ is found to give the superior fit. This model is found to give a good fit throughout the temperature range studied (100 – 300 K, lattice parameters present in ESI, Table S9). The resultant symmetry is quite curious as this phase is different from both end members at room temperature. A key feature of this symmetry is that it only has one alkali metal site, thus ruling out any possible ordering of $Cs^+$ and $Rb^+$. The lack of a phase transition in the temperature range studied is reinforced by the lack of an anomaly in the molar unit cell volume (Fig 6).

Reducing Cs content further leads to another change; for $Cs_{0.5}Rb_{1.5}SnCu_3F_{12}$ the room temperature pattern is found to be well fitted by the model previously found for $Rb_2SnCu_3F_{12}$ ($R\bar{3}$, $a$ = 14.01939(12), $c$ = 20.3565(3)). It is impossible to determine if fluorine disorder is also present in this phase as the corresponding difference between the powder patterns is minimal. On reducing temperature a great deal of peak broadening/splitting is present (Fig. 7a) and the quality of fit for this rhombohedral model diminishes. These broadenings and splittings are retained down to the lowest temperature surveyed (100 K). It is found that the low temperature pattern is best fitted by the triclinic model ($P\bar{1}$) previously derived for $Rb_2SnCu_3F_{12}$ (lattice parameters given in ESI, Table S10). For $Rb_2SnCu_3F_{12}$, there is a return to the rhombohedral phase at lowered temperature – such re-entrant behaviour is not seen for $Cs_{0.5}Rb_{1.5}SnCu_3F_{12}$. The exact transition point was once again determined by assessing the quality of fit (Fig. 7b) for both the high temperature and low temperature models. This suggests that the phase transition is at ~ 280 K, 30 K higher than that found for the similar transition in $Rb_2SnCu_3F_{12}$.

From the above (and taking into account the data for $Rb_2SnCu_3F_{12}$) a number of conclusions may be drawn about the behaviour of this solid solution. The change in molar unit cell volume, as a function of Cs content, is non-linear, and is found to be governed largely by an expansion in $ab$-plane, with very little change along the $c$-axis (Fig. 8b). This trend is opposite to that found for the variation in lattice parameters of $Cs_2SnCu_3F_{12}$ (rhombohedral phase) with temperature – in that case it is found that the $c$-axis expands significantly more than the a-axis (Fig 8c). The phase transition behaviour is summarised in Table 2. On doping Rb into $Cs_2SnCu_3F_{12}$ the temperature of the $R\bar{3}m$ to $P2_1/n$ transition increases, so that for x = 1, this transition (if it occurs) lies above ambient temperature. For x > 1, the phase behaviour changes, and the lower symmetry, $R\bar{3}$, structure of the $Rb_2SnCu_3F_{12}$ end-member occurs at ambient temperature. For both this end-member, and for the x = 1.5 composition there is a sub-ambient transition to a triclinic phase. The re-entrant behaviour seen in powdered samples of $Rb_2SnCu_3F_{12}$[10] does not occur for the x = 0.5 composition, however.

Table 2: Structural transition temperatures for $Cs_{2-x}Rb_XSnCu_3F_{12}$ (x = 0, 0.5, 1.0, 1.5, 2.0). For x = 1.0 the temperature is predicted. For x = 2.0 only the highest temperature transition is noted.

| x | Transition temperature (K) | High temperature symmetry | Low temperature symmetry |
|---|---|---|---|
| 0.0 | 170 | $R\bar{3}m$ | $P2_1/n$ |
| 0.5 | 250 | $R\bar{3}m$ | $P2_1/n$ |
| 1.0 | > 300 | ? | $P2_1/n$ |
| 1.5 | 280 | $R\bar{3}$ | $P\bar{1}$ |
| 2.0 | 250 | $R\bar{3}$ | $P\bar{1}$ |

**Magnetic studies of $Cs_2SnCu_3F_{12}$ powder**

Previous magnetic studies of $Cs_2SnCu_3F_{12}$ have focussed on single crystals. In the case of the powder there is apparently a different structural behaviour from that previously described for the single crystal. Hence the magnetic behaviour may also differ, which prompts the present study.

Measurements were performed on a sample that showed no Cu containing impurity phases via PXRD. The resultant data (Figure 9) suggests that there is a transition at 19.3 K which is qualitatively similar to that previously described for the single crystal. This behaviour is only present in well annealed samples, however; in samples which have not been thoroughly annealed (at least 9 days at 773 K), there is a secondary transition at increased temperature.

The fitting of experimental data in the range 200 - 300 K by the Curie-Weiss law (with inclusion of temperature independent term $\chi = \chi_0 + C/(T - \Theta)$) gives the values of $\chi_0 = 4.72 \times 10^{-4}$ emu/mol, Weiss temperature $\Theta = -268$ K and Curie constant $C = 0.994$. The $\chi_0$ consists of summation of diamagnetic Pascal's constants of every ion present in the chemical formula $\chi_0^{dia} \approx -2.28 \times 10^{-4}$ emu/mol[17] plus the van Vleck contribution from the open shells of $Cu^{2+}$ ions $\chi_0^{para} \approx 1.32 \times 10^{-4}$ emu/mol[18]. In contrast to these evaluations the experimentally found value of $\chi_0$ is large and positive. The significant enhancement of the temperature independent paramagnetic susceptibility can be a special feature of the strongly coupled magnetic subsystem in distorted kagome-type $Cs_2SnCu_3F_{12}$. The very large negative value of $\Theta$ suggests strong antiferromagnetic interactions in the system hampering therefore the determination of basic magnetic parameters at temperatures comparable to $\Theta$. The Curie constant is also significantly reduced as compared to that expected for $Cu^{2+}$ ions, with $g_\parallel = 2.48$ and $g_\perp = 2.10$[11]. Once again, this reduction can be a consequence of strong antiferromagnetic interactions present in the system even at room temperature. The structural phase transition in $Cs_2SnCu_3F_{12}$ is not seen in the magnetization at elevated temperatures. This means that the basic magnetic properties, i.e. exchange interaction parameters, are rather similar in both high-temperature and low-temperature phases. The title compound experiences a long range magnetic order transition at $T_N = 19.3$ K. The seemingly large frustration parameter $\Theta/T_N \sim 14$ should be considered as a rough estimation only since the quantities $\Theta$ and $T_N$ refer to structurally different phases. The magnetically ordered phase at $T < T_N$ possesses finite spontaneous magnetization whose origin tentatively is the Dzyaloshinskii – Moriya interaction. The sensitivity of magnetic parameters in both paramagnetic and magnetically ordered states to annealing signifies presence of isolated impurities and secondary phases.

**Conclusions**

Crystallographic analysis of $Cs_2SnCu_3F_{12}$ in powder form (by both synchrotron X-ray powder diffraction and neutron powder diffraction) has shown that on cooling from ambient temperature there is a structural transition from $R\bar{3}m$ to $P2_1/n$ symmetry at 170 K. This transition is different from that suggested (although not fully confirmed) from a previous single crystal study, where retention of rhombohedral symmetry was proposed[11]. Analysis of the structural properties of the solid solution $Cs_{2-x}Rb_xSnCu_3F_{12}$ (x = 0.5, 1, 1.5) shows contrasting structural behaviours across the family. The x = 0.5 composition exhibits similar behaviour to that found for $Cs_2SnCu_3F_{12}$ – a change from rhombohedral to monoclinic symmetry, but at increased temperature (250 K). The x = 1 composition displays monoclinic symmetry across the whole temperature range studied (100 – 300 K), suggesting that if the corresponding rhombohedral → monoclinic transition occurs the transition temperature is above room temperature. The x = 1.5 composition shows behaviour similar to that previously reported for $Rb_2SnCu_3F_{12}$[10], displaying a rhombohedral → triclinic phase transition at sub-ambient temperature. In this case the re-entrant behaviour seen in powdered $Rb_2SnCu_3F_{12}$ is not observed. Assessing the solid solution series as a whole the increased average size of the $A^+$ cation leads to anisotropic expansion of the unit cell, primarily occurring in the *ab*-plane; this contrasts with the thermally-induced expansion of the rhombohedral phase in $Cs_2SnCu_3F_{12}$, which occurs primarily along the *c*-axis.

Magnetically, a well-annealed powder sample of $Cs_2SnCu_3F_{12}$ shows similar behaviour to that previously described for a single crystal sample, although samples annealed for shorter time periods provide evidence for a possible second magnetic transition.

## Acknowledgements

Dr A. Daoud-Aladine is thanked for assistance with NPD measurements. The collaboration between the University of St Andrews and Moscow State University was funded by a Royal Society International Exchanges grant, in collaboration with the Russian Foundation for Basic Research (12-03-92604). LJD thanks the EPSRC for a PhD studentship via a Doctoral Training grant (EP/P505097/1). This work was carried out with the support of the Diamond Light Source, beamtime application EE7980. ANV acknowledges support of RFBR through grants 13-02-00174, 14-02-92002 and 14-02-92693.

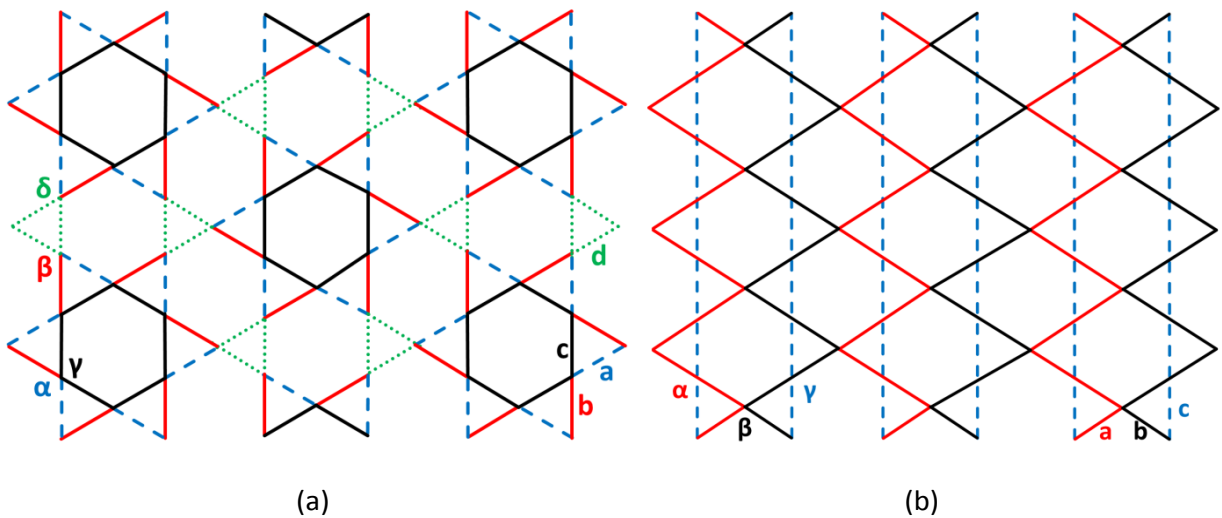

Figure 1: Schematic views of the distorted kagome lattices (Cu network only) found for the (a) $Rb_2SnCu_3F_{12}$ at room temperature ($R\bar{3}$ phase)[9] and (b) $Cs_2ZrCu_3F_{12}$ at low temperature ($P2_1/m$)[12]. For the ideal kagome lattice all triangles are strictly equilateral. For $Rb_2SnCu_3F_{12}$, the angles can vary by ~ 4 ° and the Cu – Cu bond distances vary by ~ 0.2 Å (α ≈ 121 °, β ≈ 124 °, γ ≈ 119 °, δ ≈ 114.6 °, a ≈ 3.35 Å, b ≈ 3.58 Å, c ≈ 3.53 Å, d ≈ 3.49 Å); for $Cs_2ZrCu_3F_{12}$ the corresponding deviations are smaller (α ≈ 120 °, β ≈ 119 °, γ ≈ 120 °, a ≈ 3.60 Å, b ≈ 3.61 Å, c ≈ 3.66 Å).

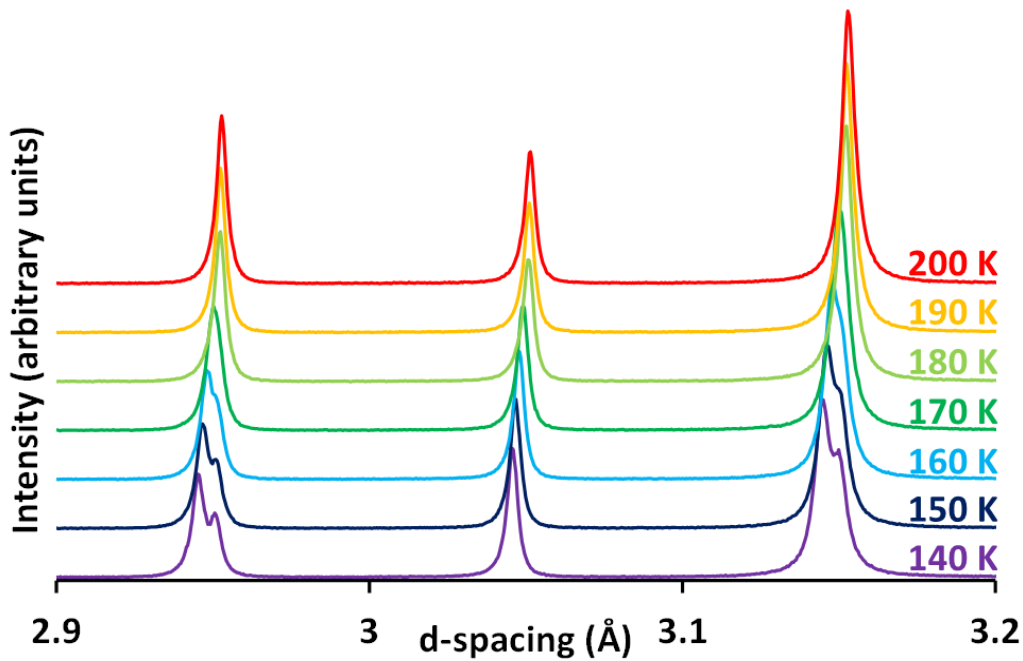

Figure 2: Portion of SXPD patterns for $Cs_2SnCu_3F_{12}$ over the temperature range 200 – 140 K. Data taken from "run 2". Note the beginnings of peak splittings (d~2.95 and 3.15 Å) at ~ 170 K.

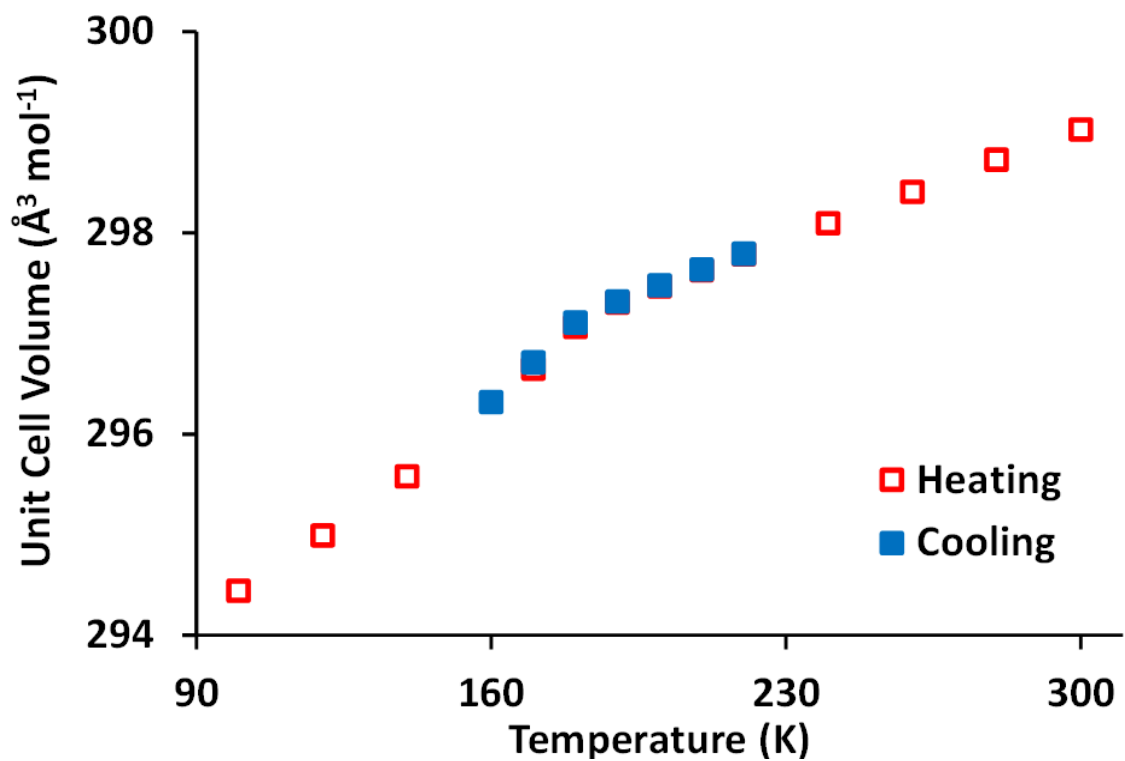

Figure 3: Unit cell volume (normalised to "per molar unit, $Cs_2SnCu_3F_{12}$") as a function of temperature for $Cs_2SnCu_3F_{12}$. Data extracted from Rietveld refinement of "run 1" SXPD data.

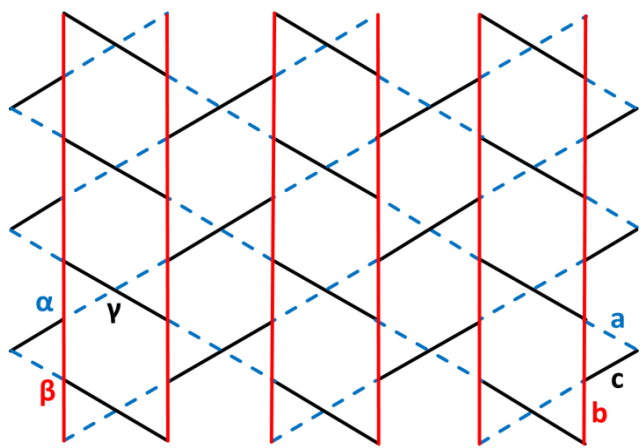

Figure 4: Schematic view of the distorted kagome lattice (Cu network only) found for $Cs_2SnCu_3F_{12}$ at low temperature ($P2_1/n$ phase)[11] .The angles vary by ~ 4 ° (α ≈ 116 °, β ≈ 123 °, γ ≈ 120 °) and the Cu – Cu bond distances vary by ~ 0.04 Å (a ≈ 3.58 Å, b ≈ 3.54 Å, c ≈ 3.55 Å)

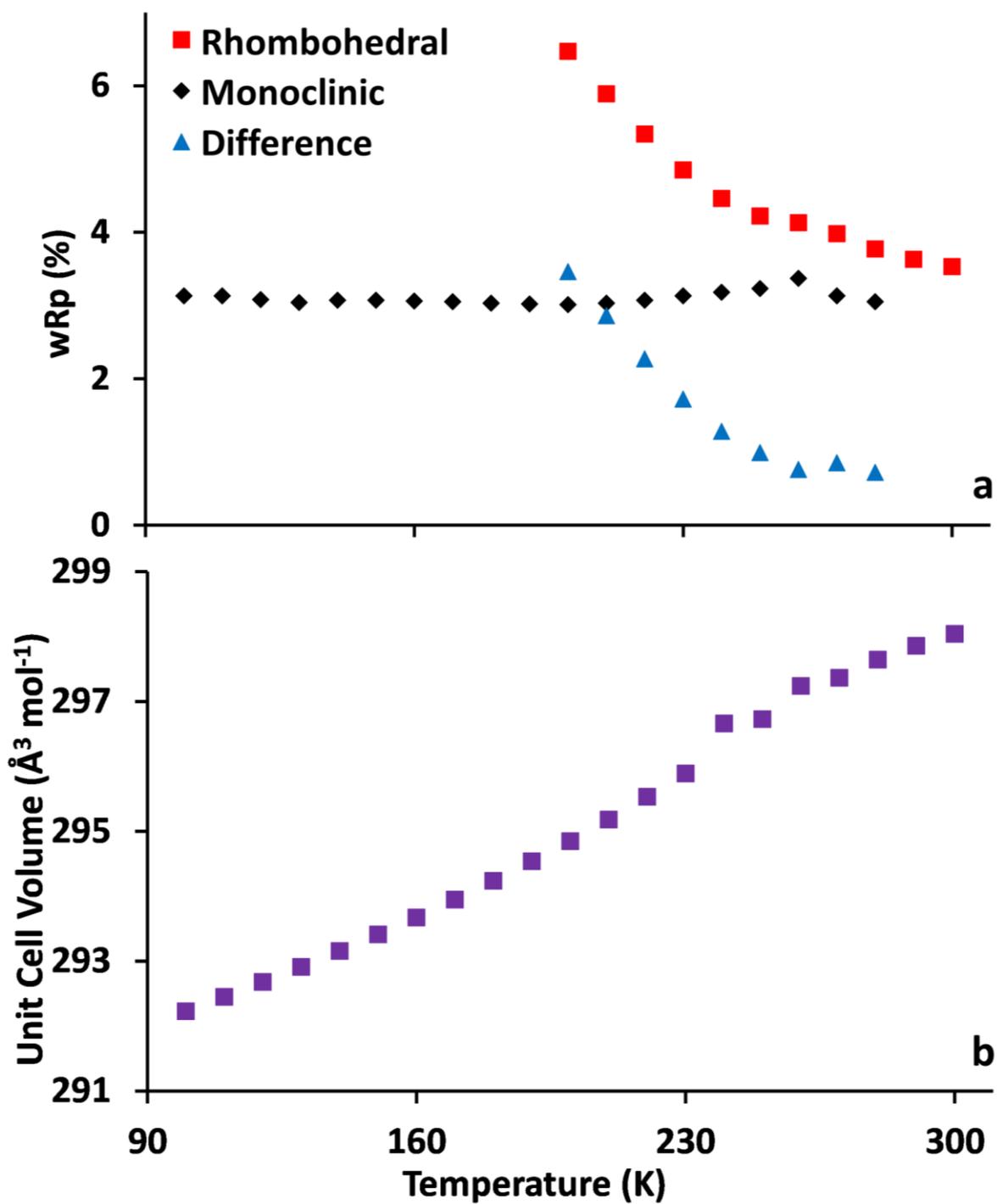

Figure 5: a, top: Quality of SXPD Rietveld fit for rhombohedral ($R\bar{3}m$) and monoclinic (P2$_1$/n) models for Cs$_{1.5}$Rb$_{0.5}$SnCu$_3$F$_{12}$ as a function of temperature. b, bottom: Volume of unit cell, per molar unit Cs$_{1.5}$Rb$_{0.5}$SnCu$_3$F$_{12}$, as a function of temperature.

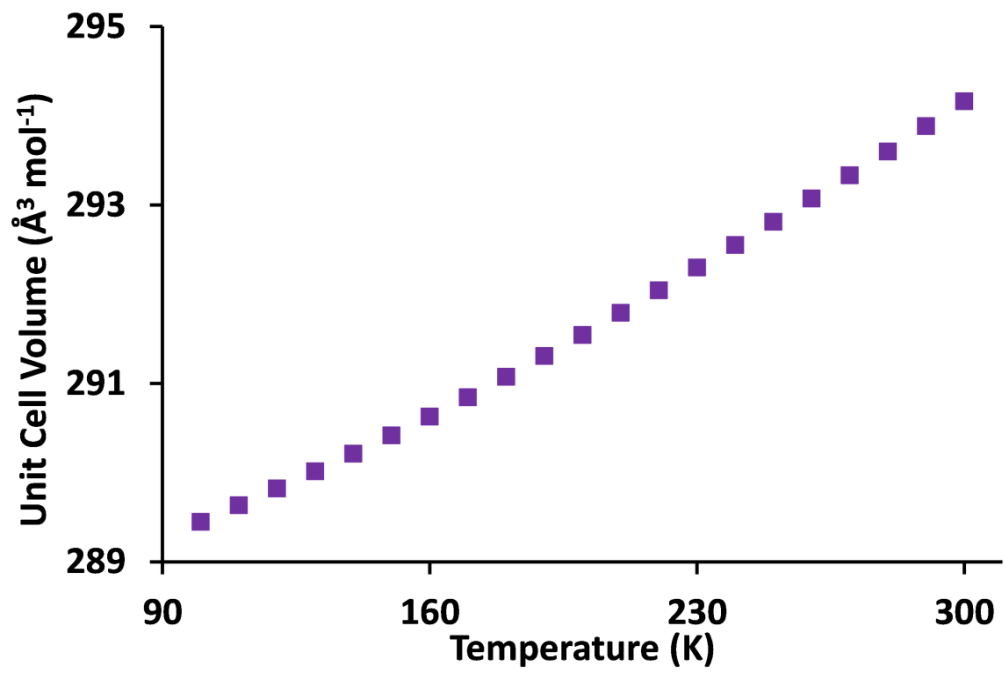

Figure 6: Molar unit cell volume as a function of temperature for $CsRbSnCu_3F_{12}$.

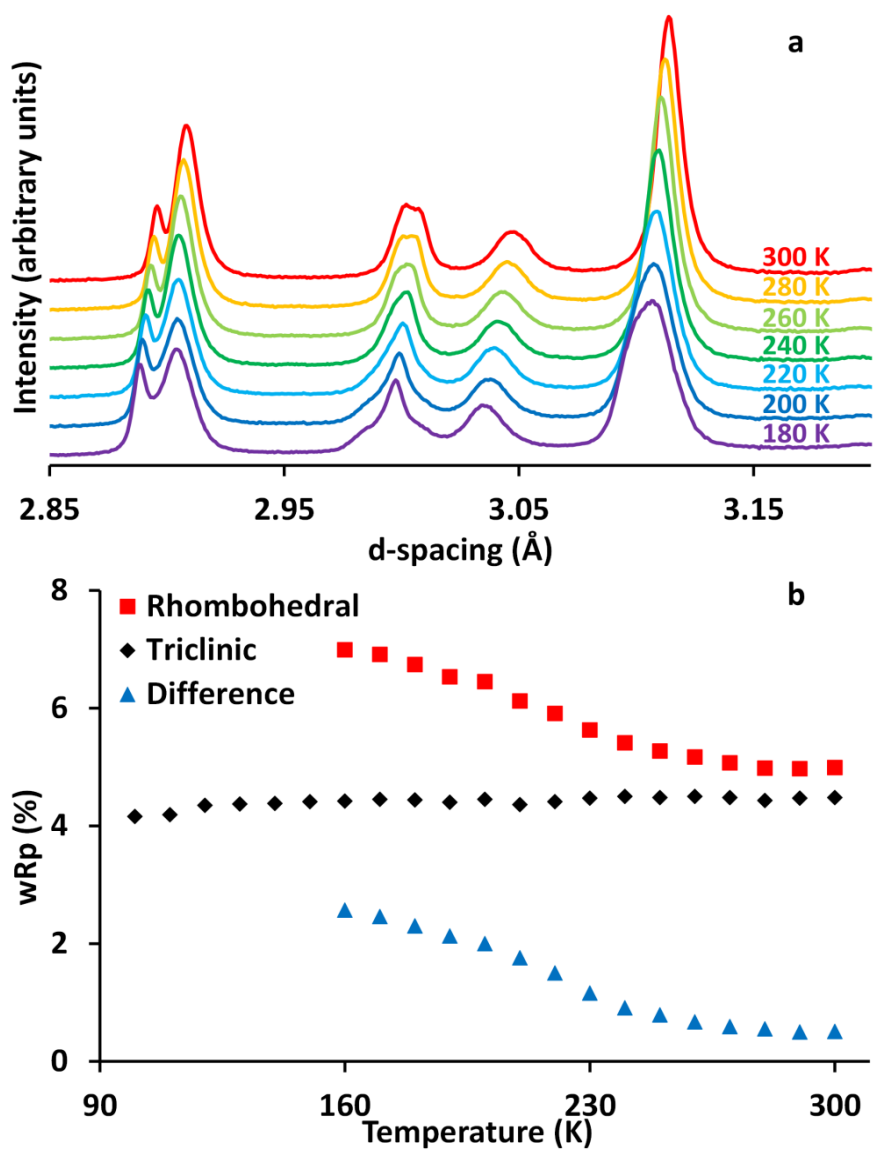

Figure 7: a: Portion of the SXPD patterns for $Cs_{0.5}Rb_{1.5}SnCu_3F_{12}$ for various temperatures. b: Quality of fit for rhombohedral (R-3) and triclinic (P-1) models for $Cs_{0.5}Rb_{1.5}SnCu_3F_{12}$ SXPD data as a function of temperature.

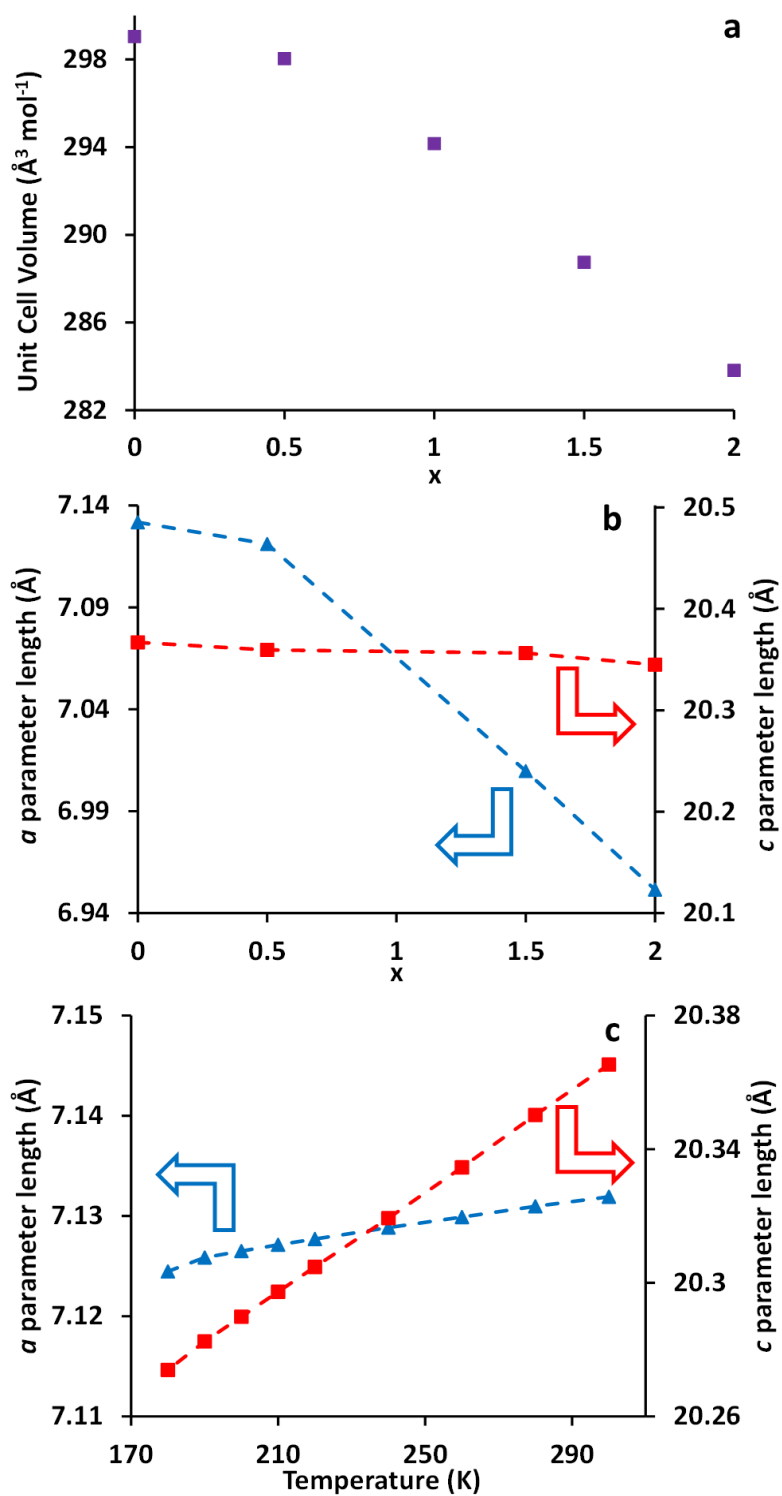

Figure 8: a: Unit cell volume per molar unit) for the series $Cs_{2-x}Rb_xSnCu_3F_{12}$ (x = 0, 0.5, 1.0, 1.5, 2.0). b: Unit cell parameters for the series $Cs_{2-x}Rb_xSnCu_3F_{12}$. c: Unit cell parameters for $Cs_2SnCu_3F_{12}$ as a function of temperature, derived from SXPD data.

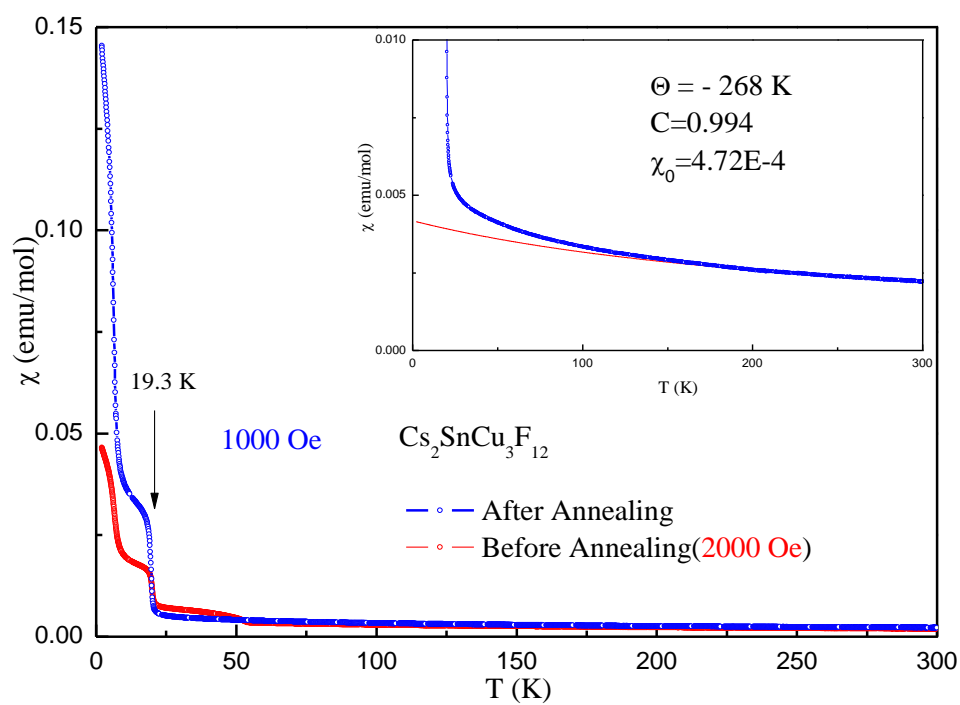

Figure 9: Magnetic susceptibility data for $Cs_2SnCu_3F_{12}$ powder; the main plot shows data recorded for two different sample treatments (note that a secondary transition near 50 K disappears on further annealing. Inset shows fitting of the experimental data by the sum of Curie-Weiss' and temperature-independent terms.